\documentclass[prl,showpacs,byrevtex,floatfix,twocolumn,superscriptaddress]{revtex4}
\usepackage{amssymb,graphicx,afterpage}
\begin{document}
\title{\boldmath Anomalous Hall effect in (In,Mn)Sb dilute magnetic semiconductor \unboldmath}
%
% authors and affiliations
%
%
\author{G. Mih\'aly}
\author{M. Csontos}
\author{S. Bord\'acs}
\author{I. K\'ezsm\'arki}
\affiliation{Department of Physics, Budapest University of
Technology and Economics, 1111 Budapest, Hungary}

\author{T. Wojtowitz}
\affiliation{Institute of Physics, Polish Academy of Sciences,
PL-02668 Warsaw, Poland}

\author{X. Liu}
\author{B. Jank\'o}
\author{J.K. Furdyna}
\affiliation{Department of Physics, University of Notre Dame, Notre
Dame, Indiana 46556, USA}
\date{\today}
%
% PACS numbers
%
\begin{abstract}
High magnetic field study of Hall resistivity in the ferromagnetic
phase of (In,Mn)Sb allows one to separate its normal and anomalous
components. We show that the anomalous Hall term is not proportional
to the magnetization, and that it even changes sign as a function of
magnetic field. We also show that the application of pressure
modifies the scattering process, but does not influence the Hall
effect. These observations suggest that the anomalous Hall effect in
(In,Mn)Sb is an intrinsic property and support the application of
the Berry phase theory for (III,Mn)V semiconductors. We propose a
phenomenological description of the anomalous Hall conductivity,
based on a field-dependent relative shift of the heavy- and
light-hole valence bands and the split-off band.
\end{abstract}
\maketitle
As promising candidates for spintronic applications (III,Mn)V dilute
magnetic semiconductors have attracted considerable attention during
the past few years \cite{1MacDonald05}. In these alloys the
Mn$^{2+}$ ions provide localized magnetic moments and valence band
holes at the same time. The presence of charge and spin degrees of
freedom and the carrier mediated nature of the ferromagnetic
coupling open a new way to the electrical control of ferromagnetism
\cite{2Ohno00}. Devices based on magnetic semiconductors represent a
novel generation of information technology where MRAM
functionalities arise simply from bulk properties
\cite{3Chiba03,4Yamanouchi04}. Moreover, the magnetic state of
diluted magnetic semiconductors (DMSs) can be quickly and
conveniently characterized by transport measurements by taking
advantage of the unusually large anomalous Hall effect (AHE). While
AHE is one of the prominent key properties of ferromagnetic
semiconductors, it represents a situation of general interest but
little firm knowledge.

In general, AHE may arise from scattering processes involving
spin-orbit coupling. In the skew scattering and side-jump models
\cite{5Sundaram99,6Yao04,7Nagaosa06} the Fermi-surface properties of
the charge carriers are important. We test this possibility
experimentally by modifying the scattering process via application
of hydrostatic pressure.

An alternative class of descriptions relates AHE to the Berry phase
acquired while the electrons propagate in spin-orbit coupled Bloch
bands. In this picture the AHE arises from near degeneracy points of
the bands
\cite{6Yao04,7Nagaosa06,8Onoda02,9Fang03,10Jungwirth02,11Murakami03}
where interband processes become relevant. Spin-orbit coupling and
the lack of inversion symmetry may result in this type of AHE even
for collinear ferromagnets \cite{10Jungwirth02}. Our high magnetic
field experiments aim to reveal an AHE contribution arising from the
relative displacement of the various spin-up and spin-down bands.

InSb has the largest spin-orbit coupling among the III-V
semiconductors, and becomes a ferromagnetic metal a when few percent
of Mn$^{2+}$ ions are inserted into the In$^+$ sites
\cite{12note1,13Wojtowitz03,14Wojtowitz04,15Csontos05,16Csontos05'}.
The band structure is well known \cite{17Vurgaftman01}, including
the parameters of the partially filled heavy hole (hh) and light
hole (lh) bands. (In,Mn)Sb is an ideal system to study AHE, and our
purpose is to distinguish between the possible AHE mechanisms
experimentally.

The In$_{0.98}$Mn$_{0.02}$Sb sample was grown by low temperature molecular beam epitaxy (MBE) in a Riber 32
R$\&$D MBE system on closely lattice matched hybrid (001) CdTe/GaAs substrates to a thickness of $230$\,nm (for
further growth details and structural characterization, see Refs. \onlinecite{13Wojtowitz03} and
\onlinecite{14Wojtowitz04}). The magnetic properties were investigated by magneto-optical Kerr-effect (MOKE)
measurements. The magnetotransport measurements were performed in a six-probe arrangement with magnetic field
perpendicular to the layer plane. For high-pressure measurements the samples were mounted in a self-clamping
cell, with kerosene as the  pressure medium.

In ferromagnetic systems the Hall-resistivity is often described as
a sum of two terms,
\begin{equation}
\rho_H=\rho_{xy}=R_0B+R_SM ,
\end{equation}
a normal Hall contribution due to the Lorentz force, plus an
anomalous Hall term ($\rho_{AH}$) that is proportional to the
magnetization $M$. Alternatively, a similar separation can be made
in the Hall-conductivity,
\begin{equation}
\sigma_H=\sigma_0B+\chi_SM .
\end{equation}
\begin{figure}[h!]
\includegraphics[width=2.5in]{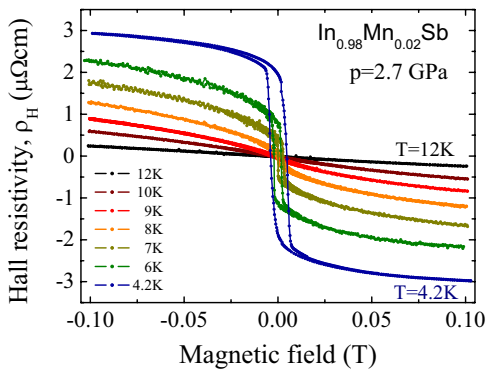}
\caption{(Color online) (a) Hall resistivity of In$_{0.98}$Mn$_{0.02}$Sb at various temperatures, up to magnetic
field $B=0.1$\,T.}
\end{figure}
Here the first term corresponds to the normal Hall effect related to the carrier concentration
($\sigma_0=R_0/\rho^2$), while the second term defines the anomalous Hall conductivity, $\sigma_{AH}$.

The analysis of the AHE in terms of $\rho_{AH}$ or $\sigma_{AH}$ is not a simple technical question. The
description by $\rho_{AH}$ implicitly assumes that the Hall signal arising from different processes is additive
in the scattering rate, 1/$\tau$. On the contrary, in the Berry phase picture the transverse current due to the
AHE is additive, and does not depend on the electron scattering that determines the longitudinal current.

In most materials - including (In,Mn)Sb - the Hall-resistivity ($\rho_{H}=\rho_{xy}$) is at least by one order
of magnitude smaller than longitudinal resistivity ($\rho_{xx}$). Consequently, the Hall conductivity defined by
Eq.~2 cannot be distinguished from the off-diagonal conductivity derived by matrix inversion from Eq.~1, as
$-\rho_{H}/(\rho_{xx}^2-\rho_{H}^2)\approx-\rho_{H}/\rho_{xx}^2=\sigma_{H}$.

In magnetic semiconductors at low magnetic fields the Lorentz term gives usually a negligible contribution
compared to the AHE, and the Hall resistivity seems to be simply proportional to the magnetization
\cite{14Wojtowitz04}. This situation is exemplified in Fig.~1 by the temperature dependence of the Hall-signal
in (In,Mn)Sb: The development of a large nonlinear contribution to $\rho_{H}(B)$ and the onset of hysteresis
loops signify the crossing of the paramagnetic-ferromagnetic phase boundary \cite{15Csontos05}.

The Hall phenomenon at high magnetic fields, however, is more complex: the anomalous Hall effect is not simply
proportional to the magnetization, i.e. the AHE coefficient - either $R_S$ or $\chi_S$ - is not a constant, but
strongly field dependent. The upper panel in Fig.~2 displays the Hall-resistivity up to $B=14$\,T, at various
pressures. Qualitatively, the two terms of Eq. (1) play different roles in the different field ranges. At low
fields, the negative contribution of the AHE overcomes the positive normal Hall term, while at high fields the
linear contribution due to the Lorentz force is dominating. Subtracting the linear term from the total Hall
signal the resulting the anomalous Hall term has a non-monotonic dependence on the magnetic field: following a
sharp peak it slowly decays, changes sign (see inset of Fig. 2), then saturates at high fields. As the
magnetization gradually increases with increasing magnetic field, it is clear that the $\rho_{AH}\propto M$
relation is not valid.
\begin{figure}[h!]
\includegraphics[width=2.9in]{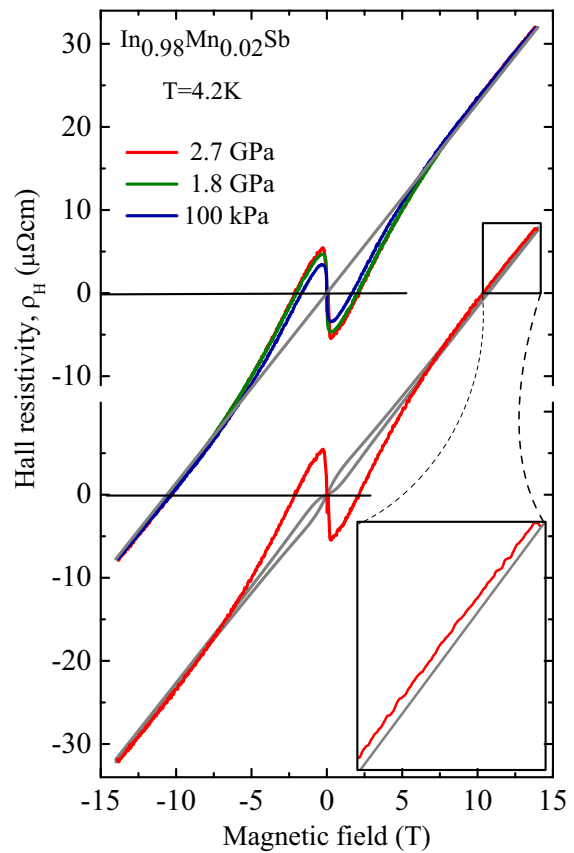}
\caption{(Color online) Upper panel: Hall resistivity of In$_{0.98}$Mn$_{0.02}$Sb at various pressures, up to
magnetic field $B=14$\,T. The gray line crossing the origin corresponds to the normal Hall effect. Lower panel:
illustration of the possible field dependence of the normal Hall resistivity due to multiband effects; the two
gray curves represent the very extreme limits for the field dependence in the normal Hall resistivity. The inset
shows the high-field behavior, where $\rho_{AH}$ saturates at a constant, positive value.}
\end{figure}

Before proceeding, we note that the separation of the normal Hall effect by subtracting a linear term - which is
often applied as a first step of the analysis - assumes that the linear field dependence of the normal Hall
resistivity is not affected by the multiband nature of the electrical conductivity. In materials where various
bands contribute to the conductivity, however, a deviation from the linear variation is expected, since the
contribution of the different types of carriers is not simply additive \cite{19Aschcroft76}. While this is true,
the characteristic field above which this effect may influence the normal Hall term is quite high, given by
$\rho_H/R_0$ which is $\sim100$\,T for (In,Mn)Sb. Another possibility is that the subband Hall contributions to
the normal Hall effect themselves are field dependent due to the change in the population of the spin-split
bands. As both the band parameters \cite{17Vurgaftman01} and the value of the exchange coupling is known
\cite{20Dietl01}, it is easy to show that in the magnetic field range of the experiments this leads to less than
$1 \%$ correction to $R_0$.

The normal Hall effect may however be field dependent for a third reason. In case of different types of carriers
any difference in the magnetic field dependence of the subband resistivities influences the relative weights of
the subband Hall contributions \cite{21note3}. While the contributions of the two subbands to the normal Hall
effect cannot be determined separately, the limiting bounds of the field dependence can easily be evaluated.
These correspond to the situations when one of the two subbands makes the dominant contribution to the Hall
effect, while the magnetoresistance arises solely either from this or from the other subband. These limiting
curves can be derived from the experimentally determined magnetoresistance. As a result, the area enclosed by
the gray curves in the lower panel of Fig.~2  represents the possible field dependence of the normal Hall
resistivity arising from the multiband nature of (In,Mn)Sb. Clearly the observed Hall signal is far beyond even
the extreme limits of the normal Hall term. Thus the strong field dependence has to be attributed to the
anomalous Hall effect. Note also that, due to the identical (positive) sign of $R_0$ in the heavy- and
light-hole bands, a negative peak cannot arise from multiband effects under any circumstances.
\begin{figure}[h!]
\includegraphics[width=2.2in]{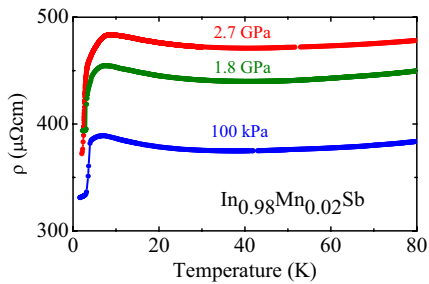}
\caption{(Color online) Temperature dependence of the resistivity at various pressures.}
\end{figure}

Next we briefly discuss the influence of hydrostatic pressure on the Hall effect. The pressure effects on the
magnetotransport have been investigated in detail previously \cite{15Csontos05,16Csontos05'}. Here we only
recall that while the number of charge carriers is not influenced by pressure (as demonstrated by the high field
behavior of the Hall resistivity shown on Fig.~2), the longitudinal resistivity is enhanced. The experimental
observations shown in Figs.~3 and 4(b) suggest that $\rho(p,T,B)$ has the form of $f(p)\rho(T,B)$, indicating
that resistivity change is due to the pressure induced enhancement of the effective mass. The diagonal
resistivity increases by more than $20\%$ for the applied pressure of $2.7$\,GPa. In contrast, the anomalous
part of the off-diagonal conductivity is independent of pressure (see Figs.~4c and 5a), indicating a
dissipationless anomalous Hall current \cite{22Lee04}.
\begin{figure}[h!]
\includegraphics[width=2.2in]{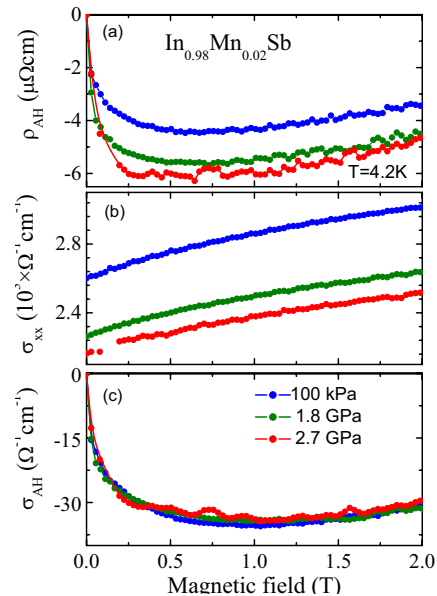}
\caption{(Color online) (a) Field dependence of the anomalous Hall resistivity at various pressures. (b) and
(c): Influence of hydrostatic pressure on the diagonal and off-diagonal terms of the conductivity tensor. In
spite of the pressure induced enhancement of the scattering process, the anomalous Hall conductivity is
pressure-independent.}
\end{figure}

The above observation is in disagreement with the extrinsic scattering picture. Simultaneously, the strong field
dependence of the AHE is also suggestive of an intrinsic mechanism, which is independent of the scattering
processes and is rather determined by the singularities in the band structure. Note also that the anomalous Hall
conductivity changes sign as a function of magnetic field (Fig. 5a), which is not expected in the scattering
models of the AHE.

Berry phase calculations of the AHE are based on the four band spherical Luttinger model of the (III,V)
semiconductors which takes into account parabolic dispersions for heavy-hole and light-hole bands in the
presence of spin-orbit coupling \cite{10Jungwirth02,11Murakami03}. Furthermore, in the ferromagnetic case an
additional term,
\begin{equation}
H_{ex}\propto J_{pd}sS\,
\end{equation}
has to be introduced into the total Hamiltonian \cite{10Jungwirth02}, which represents the exchange interaction
between the localized magnetic moments on Mn$^{2+}$ ions ($S$) and the spins of the charge-carrying holes ($s$).
This coupling results in the spin splitting of the valence bands. Jungwirth et al. have shown that if both
spin-orbit and exchange coupling ($J_{pd}$) are important, then AHE is generally nonlinear in the magnetization,
and it may have both positive and negative signs. Their numerical calculations - including the influence of the
split-off band and also the nonparabolic nature of the valence bands - gave good estimates for the magnitude of
the AHE in (Ga,Mn)As and (In,Mn)As.
\begin{figure}[h!]
\includegraphics[width=3.4in]{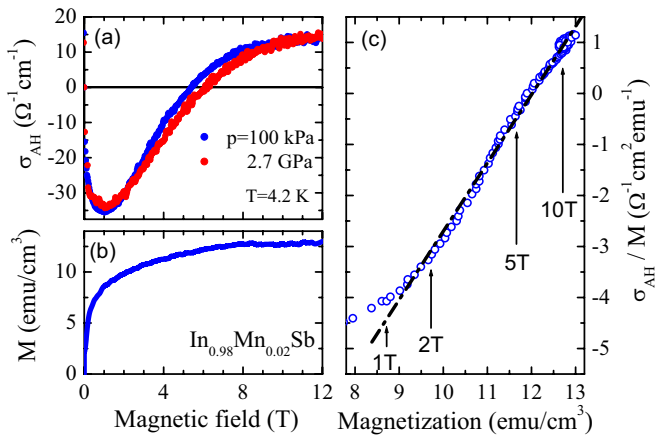}
\caption{(Color online) Comparison of the field dependence of (a) the high field anomalous conductivity and (b)
the magnetization. (c) $\sigma_{AH}/M$ as a function of $M$. The variation of $\sigma_{AH}/M$ as a function of
$M$ is clearly seen to be linear over a wide field range above about $2$\,T. The dash-dotted line corresponds to
Eq.~(4).}
\end{figure}

The Luttiger parameters of InSb \cite{17Vurgaftman01} indicate that for In$_{0.98}$Mn$_{0.02}$Sb, where the hole
concentration is $n=3\cdot10^{20}$\,cm$^{-3}$, the Fermi level is only $\sim150$\,meV away from the lower lying
split-off band. Due to the exchange splitting, the majority and minority spin bands move about $\pm25$\,meV
apart from each other. In the Berry-phase picture, the dominant contributions to AHE arise from the nearly
degenerate points of the bands located close to the Fermi energy \cite{6Yao04,7Nagaosa06,8Onoda02,9Fang03}. The
vicinity of the split-off band - even without band-crossing - may then have a significant effect on the Berry
phase acquired by the heavy and light holes, and the large shift of up- and down-spin bands may be responsible
to the measured field dependence of the AHE.

It is important to note that as the magnetic field is varied, the relative position of the bands shifts linearly
with the absolute value of the magnetization (due to the exchange origin of the splitting). Assuming that the
corresponding correction in the anomalous Hall coefficient is also linear in band shift, i.e.
$\chi_S\sim(1-\alpha|M|)$, one obtains the anomalous Hall conductivity varying as
\begin{equation}
\sigma_{AH}\propto M(1-\alpha|M|)\,
\end{equation}
Such a behavior is demonstrated in Fig.~5, where the experimentally determined $\sigma_{AH}(B)$ and the
corresponding $M(B)$ curves are plotted in the from of $\sigma_{AH}(B)/M$ versus $M$. The observed linear
variation above $B\approx2$\,T confirms the above phenomenological picture (using only one fitting parameter,
$\alpha= 0.05$\,cm$^3/$emu) \cite{23note4}.

In conclusion we showed that - in contrast to the general belief - in (In,Mn)Sb the AHE not simply proportional
to magnetization. The anomalous Hall signal (either $R_S$ or $\chi_S$) can even reverse sign before saturating
at high field. We attribute this behavior of AHE to Berry phase effects, and we propose a qualitative
description of the field-dependent AHE, where exchange splitting leads to a relative shift between the valence
bands and the nearby split-off band. The intrinsic nature of the AHE was also confirmed by high pressure
experiments: we demonstrated that the off-diagonal terms of the conductivity tensor are not influenced, while
the diagonal terms are reduced due to the pressure-induced enhancement of the scattering process.

\section*{Acknowledgement}
Several enlightening discussions with G. Zar\'and are acknowledged. We are grateful to C. Timm whose comment
initiated the clarification of multiband effects in $R_0$. This research was supported by the National Science
Foundation NIRT Award DMR 02-10519 and by the Hungarian Scientific Research Fund OTKA under Grant Nos.  F61413,
K62441, and Bolyai 00239/04.

%
% References
%

%

\begin{references}
\bibitem{12note1}The Mn doping makes the system metallic with $k_Fl\sim30$, and
ferromagnetic with Curie temperature slightly below $10$\,K (where
$k_F$ is the Fermi-wave number and $l$ is the mean fee path). The
transport and magnetic properties of (In,Mn)Sb have been
characterized in a series of recent experiments: see Refs.
\onlinecite{13Wojtowitz03,14Wojtowitz04,15Csontos05,16Csontos05'}.
\bibitem{23note4}Note that in the proposed description the pressure dependence of $J_{pd}$ [15]
rescale only the horizontal axis of the $\sigma_{AH}(B)$ curve shown
in Fig.~3 (by about $\sim3\%$, which is not resolved in the
experiment).
\end{references}
\end{document}